\documentclass[11pt,twoside]{article}
\usepackage{asp2010}

\resetcounters

\begin{document}

\title{Correcting the Astrometry of DASCH scanned plates}
\author{{M. Servillat$^1$, E. J. Los$^1$, J. E. Grindlay$^1$, S. Tang$^1$, and S. Laycock$^2$}\\
\affil{$^1$Harvard College Observatory, 60 Garden St., Cambridge, MA 02138, USA}
\affil{$^2$Department of Physics, University of Massachusetts, Lowell, MA, USA}
}

\begin{abstract}
We describe the process implemented in the DASCH pipeline which applies a reliable astrometric correction to each scanned plate. 
Our first blind astrometric fit resolves the pointing, scale and orientation of the plate in the sky using astrometry.net code.
Then we iteratively improve this solution with WCSTools \textit{imwcs}. 
Finally, we apply a 6th order polynomial fit with SCAMP to correct the image for distortions. 
During a test on 140 plates, this process has allowed us to successfully correct 79\% of the plates. 
With further refinements of the process we now reach a 95\% success rate after reprocessing all our scanned plates ($\sim$11\,000 in Nov. 2010).
We could extract a lightcurve for 2.85 times more objects than with the previous Pipeline, down to magnitude 17. The resulting median RMS error is 0.13'' for objects with mag. 8 to~17.
\end{abstract}

\section{Introduction}

The DASCH project (Digital Access to a Sky Century at Harvard) is a project that aims to digitize the $\sim525\,000$ photographic plates stored at the Harvard College Observatory that were exposed on various telescopes from 1885 to 1992 \citep{Grindlay+09}. The plates cover the whole sky and provide typically 500 to 1000 images of any object brighter than the detection limit, of typically 14 to 17 B magnitude. 

We developed a specific Pipeline \citep{Grindlay+09,Laycock+10} to process the plates and store those measurements in a database \citep[see][this volume, for the description of the Pipeline]{losadassxx}. In order to extract the long-term lightcurves (over 100 years!) of an object without confusing it with a neighboring object, we need to obtain a good astrometric solution for each plate. 

An accuracy of 1 arc second ('') or lower generally allow us to associate an object with its entry in the GSC 2.3.2 catalog uniquely, or to classify it securely as a new transient event. In practice, the scale of the plates varies from sub-arc-second to about 6'' per pixel depending on the plate series and we expect to obtain positional accuracy lower than a 3 pixels limit (radius used for cross-correlations). Distortions from the original telescope optics can have dramatic effects, with offsets of up to few arc minutes on the edges.

\section{Implementation}
We implemented a 3 step procedure that allows us to blindly find the position of the plate center and ultimately correct the distortions on the plates.

\paragraph{First guess}
It is first necessary to find the position of the plate in the sky with the right scale and orientation. 
The transcriptions of the hand-written logbooks could not reliably give this information. 
We thus need to find a solution blindly, which is possible by using pattern matching techniques. Astrometry.net procedures \citep{Lang+10} are optimized for this, and are integrated since June 2008 in our Pipeline.
In order to speed up the process, the catalog of detections is generated with SExtractor \citep{BA96} and a high threshold on a reduced image with a 16 pixels binning. 
This procedure leads to a 99.75\% success rate.

\paragraph{Refining the solution}
The first guess is too crude to efficiently match the detected objects. We thus use WCSTools \citep{Mink02} \textit{imwcs} to iteratively reach a more precise solution. The Tycho 2 catalog is used with coordinates corrected for proper motions, and generally provides a positioning of the plate with a precision of 10-20''.

\paragraph{Fitting the distortions}
For each plate, a low threshold detection of objects is performed with SExtractor. In order to correct the distortion, we select the $10\,000$ brightest sources and a reference catalog of $10\,000$ is extracted from UCAC3 \citep{Zacharias+10}. This reference catalog has indeed the best astrometric references to date (0.015 to 0.100'') as well as the best proper motions estimates, which are important on 100 year timescales. UCAC3 is filtered to remove its known biases: we keep objects with a 2MASS counterpart, and keep only reference stars with a M magnitude in the range 8-16. With those inputs, we run SCAMP \citep{Bertin06} to obtain a 6th order polynomial fit stored in the header of the plate image file for use in the following Pipeline steps. This fit was previously performed with IRAF \textit{ccmap} and the GSC catalog.

\section{Test and Results}

We performed a test on 140 plates chosen randomly from different plate series. We took care to respect the proportion of already scanned plates for each series so the results could be extrapolated to the larger number of plates now scanned. 

This resulted in 79\% of the plates correctly processed, with a mean error well below the 3~pixels limit and close or lower than 1'' (depending on the series). We note that 44\% have even better accuracy than with the previous version of the Pipeline. Distortion maps and plots for plates with different scales are shown in Figure~\ref{fig1}. 

The other 21\% of the plates showed various problems. The problems discovered during this test do not seem to be linked to SCAMP, and they generally could be solved by refining the parameters. We present examples of erratic maps in Figure~\ref{fig2}. For map A, a cloud in the sky was reported in the logbook, seen as a hole in the map. 
Plates with spotting or uneven development will produce similar holes.
For map B, the plate is possibly over-exposed and the background emission is particularly high in the center of the plate due to vignetting. By changing the detection parameters, we could recover more matches with reference stars and in both cases improve the astrometry.

For 15\% of the plates, the SCAMP correction step could not even be applied due to an insufficient number of matches with reference stars that are only locally distributed (see Figure~\ref{fig2}, C and D). This showed us that the initial astrometry is too uncertain for SCAMP to work, with scaling issues in RA and Dec.
We reviewed the code used for the second step to provide a better estimate of the astrometry before using SCAMP. 

We now reach a success rate of $\sim$95\% after reprocessing all the scanned plates to date (more than $11\,000$). We extracted photometry for a subset of $\sim$3300 plates covering one field and obtained 2.85 times more objects with a lightcurve, down to magnitude 17 (i.e. 2 magnitudes fainter than with the previous version of the Pipeline). The median RMS error on the position is about $0.13''$ for objects with magnitude 8 to 17.

\articlefigure{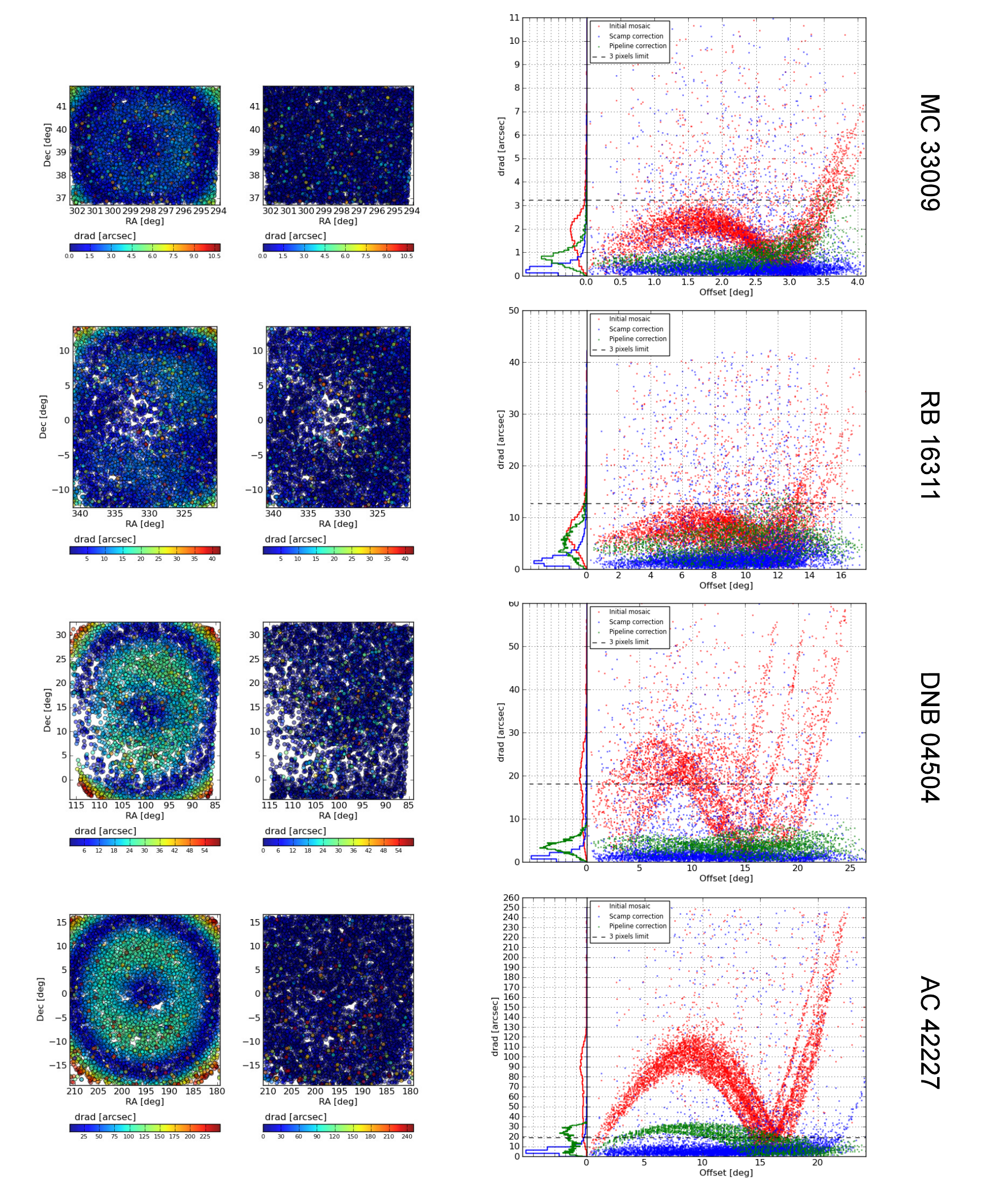}{fig1}{Distortions observed for 4 plates. \textit{Left}: Distortion maps showing each reference stars in RA/Dec with a color corresponding to the position error (drad: radial distance from reference to detection). Maps before (left) and after (right) SCAMP correction. \textit{Right}: Position error as a function of the offset from the plate center for each reference star. Red: before correction. Green: previous Pipeline. Blue: Pipeline with SCAMP. Some scattered points with high drad correspond to mismatches.}

\articlefigure[width=\textwidth]{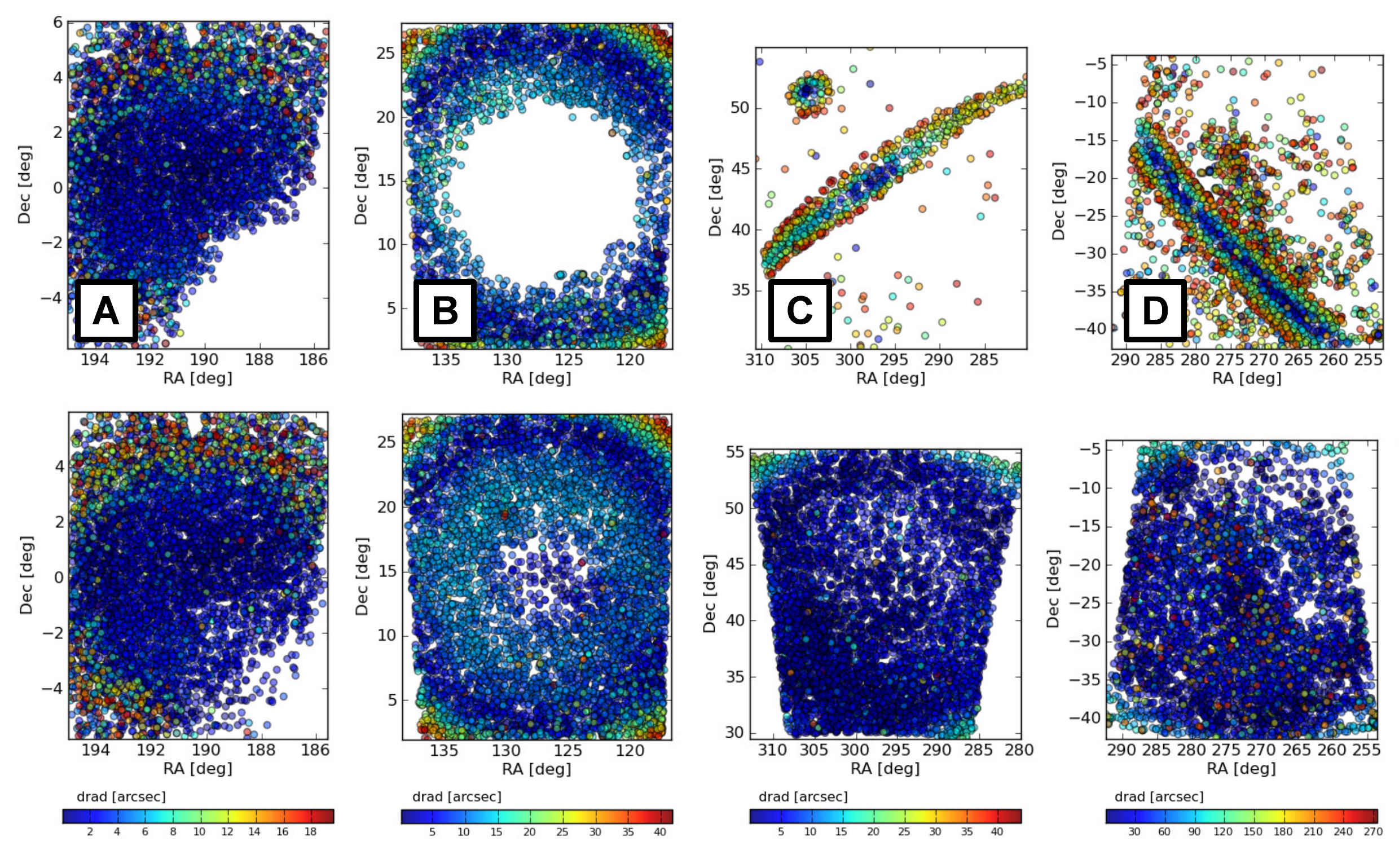}{fig2}{Erratic distortion maps before SCAMP correction. Top: results from the test on 140 plates. Bottom: latest results after refinement of the process. A:~B~71955, B:~BM~00389, C:~RH~06526, D:~AM~00538}

\acknowledgements 
DASCH is supported by NSF grants AST-0407380 and AST-0909073. MS gratefully thanks A. Doane for her comments and for sharing her knowledge on the plate collection.
\begin{center}Visit the DASCH website at http://hea-www.harvard.edu/DASCH\end{center}

\bibliography{P034}

\end{document}